\documentclass [10pt]{article}
\usepackage{amsfonts}
\usepackage{multicol}
\usepackage{setspace}
\addtocounter{page}{0}
\begin{document}

\begin{center}
\noindent
\Large{\bf The Yang-Mills Field Strength Revisited}\\

\vskip .2in

\normalsize
Samuel L. Marateck\\

\small{\textsf{\textit{Courant Institute of Mathematical Sciences, New York
    University, New York, N.Y. 10012}}}\\
\end{center}

\vskip .2in

\small{The Yang-Mills field strength incorporating a non-abelian feature is one
of the cornerstones of the standard model. Although Yang-Mills gauge theories
have been around for over fifty years, surprisingly the derivation of the
Yang-Mills field strength using classical gauge theory does not appear anywhere
in the literature. In their 1954 paper, Yang and Mills had to invent a
non-Abelian field strength to satisfy certain criteria. In Section {\bf 5} we
use Yang's gauge transformation in a heuristic derivation of the Yang-Mills
field strength.  The preceding sections cover material relating to the
derivation. Section {\bf 3} shows where Pauli in the article cited by Yang and
Mills gives an expression for the electro-magnetic field strength in terms of a
commutator. For some reason, Yang and Mills did not use this approach.}\\

\noindent
\section{\textsf{\normalsize{INTRODUCTION} }}
Although Yang-Mills (YM) gauge theory is now done using fiber-bundel theory --
see, for instance, the review articles of Daniel [1] and Marateck [2]
-- it is of interest to analyse the way it developed using classical gauge
theory.

\noindent
In their seminal paper [5], when Yang and Mills discuss the phase
factor - gauge transformation relationship, they cite Pauli's review paper
[3]. It is interesting that, although Pauli in that paper presents
the electromagnetic field strength in terms of a commutator, for whatever
reason Yang and Mills did not use the commutator to obtain the Yang-Mills (YM)
field strength -- they obtained it by generalizing the electro-magnetic field
strength. We provide an analysis of the steps required to derive the YM field
strength in this way. Also presented is a derivation of this field strength
using the commutator approach and the derivation of the YM field transformation
using a slightly different way than is traditionally done.

\noindent
\section{\textsf{\normalsize{GAUGE THEORY} }}
Weyl [4] introduced as a phase factor an exponential in which the phase
$\alpha$ is preceded by the imaginary unit $i$, e.g., $e^{+iq\alpha({\bf x})}$,
in the wave function for the wave equations (for instance, the Dirac equation
is $(i\gamma^\mu \partial_\mu - m)\psi = 0$).  It is here that Weyl correctly
formulated gauge theory as a symmetry principle from which electromagnetism
could be derived.  It had been shown that for a quantum theory of charged
particles interacting with the electromagnetic field, invariance under a gauge
transformation of the potentials required multiplication of the wave function
by the now well-know phase factor.  Yang on page 19 of his selected papers [6]
cites Weyl's gauge theory results as reported by Pauli [3] as a source for
Yang-Mills gauge theory; although Yang didn't find out until much later that
these were Weyl's results. Moreover, Pauli's article did not explicitly mention
Weyl's geometric interpretation. It was only much after Yang and Mills
published their article that Yang realized the connection between their work
and geometry. In fact, on page 74 of his selected papers [6], Yang says

\begin{quotation}
\noindent
What Mills and I were doing in 1954 was generalizing Maxwell's theory. We
knew of no geometrical meaning of Maxwell's theory, and we were not looking
in that direction.
\end{quotation}

\noindent
For the wave equations to be gauge invariant, i.e., have the same form after
the gauge transformation as before, the local phase transformation $\psi({\bf
x}) \rightarrow \psi(x)e^{+i\alpha({\bf x})}$ has to be accompanied by the
local gauge transformation

 \begin{equation} {\bf A_\mu} \rightarrow {\bf A_\mu} - q^{-1} {\bf
    \partial_{\mu}\alpha({\bf x})} \end{equation}

\noindent
This dictates that the $\partial_\mu$ in the wave equations be replaced by the
covariant derivative $\partial_\mu + iqA_\mu$ in order for the ${\bf
\partial_{\mu}\alpha({\bf x})}$ terms to cancel each other.  This pair of phase
factor- gauge transformation is not unique. Another pair that retains gauge
symmetry and results in the same covariant derivative has the $q$ included in
the phase factor, i.e., $\psi({\bf x}) \rightarrow \psi(x)e^{+iq\alpha({\bf
x})}$ paired with 

 \begin{equation}{\bf A_\mu} \rightarrow {\bf A_\mu} - {\bf \partial_{\mu}\alpha({\bf\ x})}
 \end{equation}

\noindent
The fact that this pairing is not unique is not surprising since a change in
the phase factor and gauge transformation have no physical significance.\\

\noindent
\section{\textsf{\normalsize{YANG-MILLS FIELD STRENGTH}}}

Pauli, in equation 22a of Part I of his 1941 review article [3] gives the
electromagnetic field strength in terms of a commutator\footnote{It is
presented in Pauli's equation (22a) $D_iD_k-D_kD_i = -i\epsilon f_{ik}$ where
$D_k = (\partial/\partial{x_k})-i\epsilon\phi_k$, and $\phi_k$ is the
electromagnetic potential, $\epsilon$ is the charge, and $f_{ik}=
(\partial\phi_k/\partial{x_i})-(\partial\phi_i/\partial{x_k})$ is the field
strength.}. In present-day usage it is

\begin{equation}[D_\mu, D_\nu] = i\epsilon F_{\mu \nu}\end{equation}

\noindent
where $D_\mu$ is the covariant derivative $\partial_\mu +i\epsilon A_\mu$.
Mathematically, equation (3) corresponds to the curvature (the field strength)
reflecting the effect of parallel transport of a vector around a closed path,
i.e., its holonomic behavior. If the field strength is zero, the vector will
return to its point of origin pointing in its original direction. In their
1954 paper [5] Yang and Mills do not mention this relation, although
they do cite Pauli's 1941 article [3]. They use

 \begin{equation}\psi = S\psi' \end{equation}

\noindent
where $\psi$ is a wave function and $S$ is a local isotopic spin rotation
represented by an SU(2) matrix, to obtain the gauge transformation in equation
3 of their paper

 \begin{equation} B'_\mu = S^{-1}B_\mu S + iS^{-1}(\partial_\mu S)/\epsilon \end{equation}

\noindent
where $B_\mu$ represents a field incorporating isotopic spin matrices.
They\footnote{Yang had earlier started studying this problem as a graduate
student at the University of Chicago and derived equation (5). When he returned
to this problem as a visitor at Brookhaven, he in collaboration with Mills
obtained (as we will explain) the field strength.  See page 17 in Yang's
selected papers, Ref. 3.} then define the field strength as

 \begin{equation}F_{\mu \nu} = (\partial_{\nu}B_{\mu} -\partial_{\mu}B_{\nu}) +i\epsilon
     (B_\mu B_\nu - B_\nu B_\mu) \label{eq:F} \end{equation}

\noindent
They didn't know at that time that this corresponds to Cartan's second
 structural equation which in differential geometry notation is ${\bf \Omega =
 dA + [A, A]}$, where $A$ is a connection on a principal fiber bundle.\\

\noindent
On page 19 of his selected papers [6], Yang states

\begin{quotation}
\noindent
Starting from [$\psi = S\psi'$ and $S(\partial_\mu -i\epsilon B'_\mu) \psi' =
(\partial_\mu -i\epsilon B_\mu)\psi$] it was easy to get [our equation (5)].
Then I tried to define the field strength  $F_{\mu \nu}$ by $F_{\mu \nu} = \partial_{\nu}B_{\mu}
-\partial_{\mu}B_{\nu}$ which was a "natural" generalization of
electromagnetism. 
\end{quotation}

\noindent
Yang returned to this work when he collaborated with Mills when they shared an
office at Brookhaven. They published their results in their 1954 paper [5].
There they introduce equation (6) (their equation 4) by saying

\begin{quotation}
\noindent
In analogy to the procedure of obtaining gauge invariant field strengths in the
electromagnetic case, we define now
\begin{center}
 (4)$F_{\mu \nu}=(\partial_{\nu}B_{\mu} -
\partial_{\mu}B_{\nu})+i\epsilon (B_\mu B_\nu-B_\nu B_\mu)$
\end{center}

\noindent
One easily shows from [$B'_\mu = S^{-1}B_\mu S + iS^{-1}(\partial_\mu
S)/\epsilon $] that
\begin{center}
(5)$F'_{\mu \nu} = S^{-1}F_{\mu \nu}S$ 
\end{center}

\noindent
under an isotopic gauge transformation. Other simple functions of $B$ than (4)
do not lead to such a simple transformation property.
\end{quotation}

\noindent
In section {\bf 5}, we show that by substituting our equation (5) into the
electromagnetic field strength $F'_{\mu \nu} = \partial_{\nu}B'_{\mu}
-\partial_{\mu}B'_{\nu}$ dictates adding a non-electromagnetic term, i.e., the
non-Abelian term, so that their equation (5), the similarity transformation, is
satisfied.

\noindent
Using the Yang-Mills covariant derivative $(\partial_\mu - i\epsilon B_\mu)$
let's see how the Yang-Mills field strength is obtained from the commutator

\begin{center}
$[D_\mu, D_\nu] = (\partial_\mu - i\epsilon B_\mu)(\partial_\nu -
i\epsilon B_\nu) -$ \end{center}
 \begin{equation} 
(\partial_\nu - i\epsilon B_\nu)(\partial_\mu - i\epsilon
B_\mu) \end{equation}

\noindent
operating on the wave function $\psi$.  Note that $-\partial_\mu (B_\nu \psi)
= -(\partial_\mu B_\nu) \psi - B_\nu \partial_\mu \psi$ and $\partial_\nu
(B_\mu \psi) = (\partial_\nu B_\mu) \psi + B_\mu \partial_\nu \psi$.  So we get
a needed $- B_\nu \partial_\mu$ and a $B_\mu \partial_\nu$ term to cancel
$B_\nu \partial_\mu$ and $-B_\mu \partial_\nu$ respectively. Thus expanding
(7) we get

 \begin{center}
$\partial_\mu \partial_\nu - i\epsilon \partial_\mu B_\nu - i\epsilon
B_\mu \partial_\nu - i\epsilon B_\nu \partial_\mu - \epsilon^2 B_\mu B_\nu -
\partial_\nu \partial_\mu $ \end{center}
 \begin{equation} 
+ i\epsilon \partial_\nu B_\mu + i\epsilon B_\nu
\partial_\mu + i\epsilon B_\mu \partial_\nu + \epsilon^2 B_\nu B_\mu 
 \end{equation}

\noindent
which reduces to $i\epsilon (\partial_\nu B_\mu - \partial_\mu B_\nu) -
\epsilon^2 [B_\mu, B_\nu]$ or $[D_\mu, D_\nu] = i\epsilon F_{\mu \nu}$

\noindent
\section{\textsf{\normalsize{THE FIELD TRANSFORMATION}}}
We present a detailed derivation of the gauge transformation by using the
transformation

 \begin{equation}\psi' = S\psi \end{equation}

\noindent
instead of the traditional $\psi = S\psi'$, i.e., the one Yang and Mills used. 
In order to obtain the gauge transformation in equation 3 of the Yang and
Mills paper

\begin{equation}
B'_\mu = S^{-1}B_\mu S + iS^{-1}(\partial_\mu S)/\epsilon 
\end{equation}

\noindent
requires you to use\footnote{The following can be obtained by differentiating
$S^{-1}S = I$} $\partial_\mu S^{-1} = - S^{-1} (\partial_\mu S) S^{-1}$. Thus,
the approach indicated by equation (9) is marginally more
straight-forward since it doesn't require differentiating the inverse of a
matrix. You must, however, perform this differentiation when deriving the
field strength.

\noindent
The covariant derivative, $D_\mu = \partial_\mu -i\epsilon B_\mu$,
transforms the same way as $\psi$ does

 \begin{equation}D'\psi' = SD\psi \end{equation}

\noindent
The left-hand side of equation (11) becomes

\begin{equation}(\partial_\mu -i\epsilon B'_\mu)S\psi= (\partial_\mu S)\psi + 
S\partial_\mu \psi -i\epsilon B'_\mu S \psi \end{equation}

\noindent
But (12) equals $S\partial_\mu\psi -i\epsilon SB_\mu \psi.$ Cancelling
$S\partial_\mu\psi$ on both sides we get,

 \begin{equation} (\partial_\mu S) \psi -i\epsilon B'_\mu S\psi = -i\epsilon SB_\mu \psi
\end{equation} or

 \begin{equation} B'_\mu S = SB_\mu + (\partial_\mu S)/(i\epsilon)  \end{equation} thus 

 \begin{equation} B'_\mu = SB_\mu S^{-1} -i(\partial_\mu S)S^{-1}/\epsilon
\end{equation}

\noindent
We will use $S = e^{i{\bf \alpha(x) \cdot \sigma}}$. So for $\alpha$
infintesimal, $S = 1 + i\alpha \cdot \sigma$ which produces

\begin{center}
$B'_\mu = (1 + i\alpha \cdot \sigma)B_\mu (1 - i\alpha \cdot \sigma ) $
\end{center}
\begin{equation} 
- i(1/\epsilon) \partial_\mu (1 + i\alpha \cdot \sigma ) (1 - i\alpha \cdot
\sigma) \end{equation}

\noindent
Remembering that $(a \cdot \sigma)(b \cdot \sigma) = a \cdot b + i\sigma \cdot
(a \times b)$, setting $B_\mu = \sigma \cdot b_\mu$, and since $\alpha$ is
infintessimal, dropping terms of order $\alpha^2$, we get

\begin{center}
$b'_\mu \cdot \sigma = b_\mu \cdot \sigma$\end{center}\begin{equation} 
+ i[(\alpha \cdot \sigma)(b_\mu
  \cdot \sigma), (b_\mu \cdot \sigma)(\alpha \cdot \sigma)] + (1/\epsilon)
\partial_\mu (\alpha \cdot \sigma ) \end{equation}and finally
\begin{equation} 
b'_\mu = b_\mu + 2(b_\mu \times \alpha)
+ (1/\epsilon) \partial_\mu
\alpha \end{equation}

\noindent
which (because our S is the inverse of Yang-Mills' S) is equation 10
in the Yang-Mills paper [5]. \\

\noindent
\section{\textsf{\normalsize{FINDING THE FIELD STRENGTH} }}

\noindent
We reconstruct how one can go about determining the field strength. Since

\begin{equation}F'_{\mu \nu} = S^{-1}F_{\mu \nu}S\end{equation}

\noindent
let's start off with the electromagnetic-like field strength in the primed system

\begin{equation}F'_{\mu \nu} = \partial_{\nu}B'_{\mu} -\partial_{\mu}B'_{\nu}
\label{eq:F} \end{equation}

\noindent
and express it in terms of the non-primed system fields. We
calculate $\partial_{\nu}B'_{\mu}$ from $B'_\mu = S^{-1}B_\mu S +
iS^{-1}(\partial_\mu S)/\epsilon$, equation (5), obtaining

\begin{center}$\partial_{\nu}B'_{\mu}=-S^{-1}(\partial_{\nu}S)S^{-1}B_{\mu}S+
S^{-1}(\partial_{\nu}B_{\mu})S+S^{-1}B_{\mu}\partial_{\nu}S$
+\end{center}
\begin{equation}
i/\epsilon[-S^{-1}(\partial_{\nu}S)S^{-1}\partial_{\mu}S+
S^{-1}\partial_{\nu}\partial_{\mu}S]
\end{equation}

\noindent
So

\begin{center}
$\partial_{\nu}B'_{\mu} -\partial_{\mu}B'_{\nu} = -S^{-1}
[(\partial_{\nu}S)S^{-1}B_{\mu}- (\partial_{\mu}S)S^{-1}B_{\nu}]S$
\end{center}
\begin{center}
$+S^{-1}[\partial_{\nu}B_{\mu}- \partial_{\mu}B_{\nu}]S+
S^{-1}[B_{\mu}\partial_{\nu}- B_{\nu}\partial_{\mu}]S+$
\end{center}
\begin{equation}i/\epsilon[-S^{-1}(\partial_{\nu}S)S^{-1}\partial_{\mu}S +
S^{-1}(\partial_{\mu}S)S^{-1}\partial_{\nu}S]\end{equation}

\noindent
We see that the $+S^{-1}[\partial_{\nu}B_{\mu}- \partial_{\mu}B_{\nu}]S$ term
satisfies equation (19) if the field strength only had the electromagnetic-like
contribution. The other terms must either represent the transformed
non-electromagnetic-like part of $F_{\mu \nu}$ or be cancelled by adding the
non-electromagnetic terms to equation (20). Since $S$ is only used for the
transformation, it should not appear in the expression for $F_{\mu \nu}$.  

\noindent
The $i/\epsilon$ term in equations (22) dictates that a term multiplied by
$i\epsilon$ be added to equation (20).  Since $S^{-1}(\partial_{\mu}S)$ and
$S^{-1}\partial_{\nu}S$ appear in the expressions for $B'_\mu$ and $B'_\nu$
respectively, the product of $S^{-1}(\partial_{\mu}S)$ and
$S^{-1}\partial_{\nu}S$ that appears in the last term of equation (22) suggests
that we should start our quest to eliminate extra terms in equation
(20) by adding $i\epsilon B'_\mu B'_\nu$ to that equation. This product gives

\begin{center}
$i\epsilon B'_\mu B'_\nu = i\epsilon[S^{-1}B_\mu S +
iS^{-1}(\partial_\mu S)/\epsilon]*$
\end{center}
\begin{equation}
[ S^{-1}B_\nu S + iS^{-1}(\partial_\nu S)/\epsilon]\end{equation}

\noindent
which equals
\\
\\
\begin{center}
$i\epsilon S^{-1} B_\mu B_\nu S -i/\epsilon S^{-1}(\partial_\mu S)S^{-1}
\partial_\nu S -$ 
\end{center}
\begin{equation}
S^{-1} B_\mu \partial_\nu S -  S^{-1}(\partial_\mu S)
S^{-1}B_\nu S\end{equation}

\noindent
All but the first term (which represents the transformation of $i\epsilon B_\mu
B_\nu$) cancel components of the extraneous terms in equation (22). And
$i\epsilon (B'_\mu B'_\nu - B'_\nu B'_\mu)$ cancels all of the extraneous terms
except the transformation of $i\epsilon(B_\mu B_\nu - B_\nu B_\mu)$.

\noindent
After performing the cancellation, we get

\begin{center}
$ \partial_{\nu}B'_{\mu} -\partial_{\mu}B'_{\nu} +i\epsilon(B'_\mu B'_\nu -
B'_\nu B'_\mu)
 = $
\end{center}
\begin{equation}
S^{-1}[\partial_{\nu}B_{\mu} -\partial_{\mu}B_{\nu} +i\epsilon(B_\mu B_\nu
- B_\nu B_\mu)]S\end{equation}

\noindent
which satisfies equation (19).

\noindent
Using equation (15) instead of equation (10), the equation Yang and Mills
used, does not simplify the generation of the Yang-Mills field strength.

\noindent
As we know today, the commutator part of the YM field strength represents the
interaction of the quanta of the $B$ field, which is due to the isospin they
carry. 

\noindent
\section{\textsf{\normalsize{CONCLUDING REMARK.}}}
The conserved current of electromagnetism is associated with the U(1) gauge
group which has the photon as its gauge particle. Because Yang and Mills worked
with an SU(2) symmetry, their theory predicted three gauge particles. However,
since the gauge invariance is local, there can be no mass term in the YM
Lagrangian, so the gauge particles predicted by the YM theory have zero mass.
This was corrected by the electroweak SU(2) x U(1) theory which incorporates YM
non-abelian gauge theory. It predicts the existence of four gauge bosons: the
three massive ones, $W^{\pm}$ and $Z^0$, and the photon.

\noindent
\section*{\textsf{\normalsize{ACKNOWLEDGEMENTS}}}
\addcontentsline{toc}{section}{\textsf{\bf ACKNOWLEDGEMENTS}}
\noindent
The author thanks J. D. Jackson for his useful comments.

\noindent
\section*{\textsf{\normalsize{REFERENCES}}}
\noindent
\small{
\noindent
[1] Daniel, M. and Viallet, C.M., Rev. Mod. Physics. {\bf 52} 175 (1980).\\
\noindent
[2] Marateck, Samuel L., Notic. Amer. Math. Soc. {\bf 53} 744 (2006);
Math. Advance in Translation, Chinese Acad. of Sciences (Chinese Translation),
{\bf 2} 97, (2009).\\
\noindent
[3] Pauli, W., Rev. Mod. Physics. {\bf 13} 203 (1941).\\
\noindent
[4] Weyl, Hermann, Zeit. f. Physic. {\bf 330} 56 (1929).\\
\noindent
[5] Yang, C. N. and Mills, R. L., Phys. Rev. {\bf 96} 191 (1954).\\
\noindent
[6] Yang, C.N., Selected Papers (1945-1980) With Commentary, World
Scientific (2005).
\end{document}